\documentclass{article}
\usepackage{multirow}
\usepackage{tabularx}
\usepackage{graphicx}
\usepackage[table,svgnames]{xcolor}\usepackage{array}
\usepackage{hhline}
\newcolumntype{C}[1]{>{\centering\arraybackslash}p{#1}}
\usepackage{amssymb,amsfonts,amsmath,mathrsfs}
\usepackage{color}
\usepackage{colortbl}
\usepackage[normalem]{ulem}
\usepackage{caption}
\usepackage{cellspace}
\usepackage{etoolbox}
\usepackage{url}
\usepackage{cleveref}

      \title{Assessing the accuracy of direct-coupling analysis for RNA contact prediction}
        \author{%
                Francesca Cuturello\,$^{1}$,
                Guido Tiana\,$^{2}$,
                Giovanni Bussi \,$^{1}$%
\footnote{To whom correspondence should be addressed.
Email: bussi@sissa.it
}}
\date{
\small
        $^{1}$
        Scuola Internazionale Superiore di Studi Avanzati, International School for Advanced Studies,
        via Bonomea 265, 34136 Trieste, Italy \\
        $^{2}$
        Center for Complexity and Biosystems and Department of Physics, Universit\`{a} degli Studi di Milano and INFN, via Celoria 16, 20133 Milano, Italy
}

\begin{document}
\maketitle

\begin{abstract}
Many non-coding RNAs are known to play a role in the cell directly linked to their structure.
Structure prediction based on the sole sequence is however a challenging task.
On the other hand, thanks to the low cost of sequencing technologies,
a very large number of homologous sequences are becoming available for many RNA families.
In the protein community, it has emerged in the last decade the idea of exploiting the covariance of
mutations within a family to predict the protein structure using the direct-coupling-analysis (DCA) method.
The application of DCA to RNA systems has been limited so far.
We here perform an assessment of the DCA method on 17 riboswitch families,
comparing it with the commonly used mutual information analysis and with state-of-the-art R-scape covariance method.
We also compare different flavors of DCA, including mean-field, pseudo-likelihood, and a proposed stochastic
procedure (Boltzmann learning) for solving exactly the DCA inverse problem.
Boltzmann learning outperforms the other methods in predicting contacts
observed in high resolution crystal structures.
\end{abstract}

\sloppy
	\section{Introduction}
	The number of non-coding RNAs with a known functional role has steadily increased in the last years \cite{morris2014rise,hon2017atlas}.
	For a large fraction of them, their function has been suggested to be directly related to their structure \cite{smith2013widespread}.
	For paradigmatic cases such as
	ribozymes \cite{doherty2000ribozyme}, that catalyze chemical reactions, and
	riboswitches \cite{serganov2013decade}, whose aptamer domain has evolved in order to specifically bind physiological metabolites,
	a well-defined three-dimensional structure is required for function.
	Secondary structure can be inferred using thermodynamic models \cite{mathews2016rna}, often used in combination with
	chemical probing data \cite{weeks2010advances}.
	Tertiary structure is usually determined using more complex
	techniques based on nuclear magnetic resonance \cite{rinnenthal2011mapping} or X-ray diffraction \cite{westhof2015twenty}.
	Predicting RNA tertiary structure from sequence alone is still very difficult,
	as it can be seen by the relatively poor predictive performances of molecular dynamics simulations \cite{sponer2018rna}
	and knowledge-based potentials \cite{miao2017rna}.
	The low cost of sequencing techniques, however, lead to the accumulation of a vast number
	of sequence data for many homologous RNA families \cite{nawrocki2014rfam}.
	Covariance of aligned homologous sequences has been traditionally used to help or validate
	three-dimensional structural modeling (see, \emph{e.g.}, \cite{michel1990modelling} and \cite{costa1997rules} for early examples).
	Systematic approaches based on mutual information analysis \cite{eddy1994rna}
        and related methods \cite{pang2005prediction}
	are now routinely used to construct covariance models and score putative contacts.
        Recently, a G-test-based statistical procedure called R-scape has been shown to be more robust
        than plain mutual information analysis for RNA systems \cite{rivas2017statistical}.
	In the last years, in the protein community it has emerged the idea of using so-called direct coupling analysis (DCA)
	in order to construct a probabilistic model capable to generate the correlations observed in the
	analyzed sequences \cite{morcos2011direct,marks2011protein,nguyen2017inverse,cocco2018inverse}: strong direct couplings in the model
	indicate spatial proximity.
	The solution of the corresponding inverse model is usually found
	in the so-called mean-field approximation \cite{morcos2011direct}, that is strongly
correlated with the sparse inverse covariance approach \cite{jones2011psicov}. A further improvement in the level of approximation of the inferred solution is reached when maximizing the conditional likelihood (or \emph{pseudo-likelihood}), which is a consistent estimator of the full likelihood but involves a tractable maximization \cite{ekeberg2013improved}
and is considered as the state-of-the-art method for protein sequences.

	%
	%
	%
	%
	Whereas covariance methods have been applied to RNA systems since a long time, the application of DCA to RNA structure prediction has so far been limited.
	The coevolution of bases in RNA fragments with known structure has been investigated \cite{dutheil2010base},
	observing strong correlations in Watson-Crick (WC) pairs and much weaker correlations in non-WC pairs.
	DCA has been first applied to RNA in two pioneering works,
	using either the mean-field approximation \cite{de2015direct}
	or a pseudo-likelihood maximization \cite{weinreb20163d}.
	A later work also used the mean-field approximation to infer contacts \cite{wang2017optimization}.
	The mentioned applications of DCA to RNA structure prediction focused on the prediction of RNA three-dimensional
	structure based on the combination of DCA with some underlying coarse-grain model \cite{de2015direct,weinreb20163d,wang2017optimization}.
	However, the performance of the DCA alone is difficult to assess from these works, since the reported results largely depend on the accuracy of the utilized coarse-grain models.
	In addition, within the DCA procedure there are a number of subtle arbitrary choices that might significantly affect
	the result, including the 
	choice of a suitable sequence-alignment algorithm and the identification of the correct threshold for contact prediction.
Due to the relatively weak degree of coevolution in RNA \cite{dutheil2010base} as compared, for instance, to proteins,
a careful analysis of the different methods that can be used to quantify it is particularly urgent. 

	In this paper, we report a systematic analysis of the performance of DCA methods for 17 riboswitch families chosen among those for which at least one high-resolution crystallographic structure is available. 
Riboswitches are ubiquitous in bacteria and thus show a significant degree of sequence heterogeneity within each family.
	A stochastic procedure based on Boltzmann learning for solving exactly the DCA inverse problem
is introduced and compared with the mean-field solution
and the pseudo-likelihood maximization approach, as well as with mutual information and R-scape method.
A rigorous cross-validation procedure that allows to find a portable threshold to identify predicted contacts is also introduced.
Whereas Boltzmann learning is usually considered as a numerically unfeasible procedure in DCA, we here
show that it can be effectively used to infer parameters that reproduce correctly the statistical properties of the analyzed alignments and
that correlate with experimental contacts better than those predicted using alternative approximations.

	\section{Materials and Methods}
	In order to conduct a proper analysis of nucleotide co-evolution, homologous RNA sequences need to be aligned through a process named multiple sequence alignment (MSA).
	A number of different algorithms have been proposed to this aim.
	The results of any co-evolutionary analysis will depend on this initial step.
	We here tested two commonly used MSA algorithms, namely those implemented in $\textit{ClustalW}$ \cite{thompson1994clustal} and $\textit{Infernal}$ \cite{nawrocki2013infernal}.

MSAs are matrices $\{\sigma^b\}_{b=1}^B$ of $B$ homologous RNA sequences that have been aligned through insertion of gaps to have a common length $N$, so that each sequence can be represented as $\sigma^b=\{\sigma^b_1 ,...,\sigma^b_N\}$.
	Vector $\sigma$ has entries from a $q=5$ letters alphabet $\{A,U,C,G,-\}$ coding for nucleotide type, where $-$ represents a gap.
	$F_i(\sigma)$ denotes the empirical frequency of nucleotide $\sigma$ at position $i$ and $F_{ij}(\sigma ,\tau)$ the frequency of co-occurence of nucleotides $\sigma$ and $\tau$ at positions $i$ and $j$, respectively:
	\begin{align}
		\label{1}
		F_i (\sigma )&=\frac{1}{B} \sum_{b=1} ^B \delta(\sigma_i ^b ,\sigma ) \\
		\label{2}
		F_{ij} (\sigma,\tau)&=\frac{1}{B} \sum_{b=1} ^B \delta(\sigma_i ^b ,\sigma ) \delta(\sigma_j ^b ,\tau )
	\end{align}
	Here $\delta$ is the Kronecker symbol (which equals one if the two arguments coincide and zero elsewhere) and $\sigma_k^b$ is the nucleotide located at position $k$ in the $b$-th sequence of the MSA.
	In order to reduce the effect of possible sampling biases in the MSA we  adopt the reweighting scheme as in \cite{de2015direct} with sequences similarity threshold 0.9. However, we did not find significant difference
in test cases where the reweighting scheme was omitted (Supporting Information, Table 5).

The idea of DCA is to construct a probability distribution
that can generate an ensemble of sequences compatible with the available ones.
The probability distribution that maximizes the entropy, thus minimizing the amount of information,
among those compatible with the empirical frequencies has the following form:
        \begin{equation}
                \label{dca}
                P(\{\sigma\})=\frac{1}{Z}\exp\left( \sum_i h_i (\sigma_i )+\sum_{i j}J_{i j}(\sigma_i , \sigma_j )\right)
        \end{equation}
The frequencies of nucleotides and co-occurence of nucleotides corresponding to the model,
$f_{i}(\sigma)$ and $f_{ij}(\sigma,\tau)$, should coincide with the frequencies observed in the MSA, $F_{i}(\sigma)$ and
$F_{ij}(\sigma,\tau)$.
The coupling matrix $J$ only contains direct interactions and is free of indirect correlations, hence the name direct couplings.
In the following we discuss the technical details associated to the Boltzmann learning procedure introduced here.
For a more general introduction to DCA and to the other methods to perform DCA (mean field and pseudolikelihood) see Supplementary Methods.

	\subsection{Maximum likelihood and Boltzmann learning}
	Given a set of independent equilibrium configurations $\{\sigma^b\}_{b=1}^B$ of the model (Eq.~\ref{dca}) such that
	$
		P(\sigma)=\prod_{b=1} ^B P (\sigma^b )
	$,
	a statistical approach to infer parameters $\{h,J\}$ is to let them maximize the likelihood, i.e. the probability of generating the data set for a given set of parameters \cite{ekeberg2013improved}.
This can be equivalently done minimizing the negative log likelihood divided by the effective number of sequences:
	\begin{equation}
		l=-\frac{1}{B}\sum _{b=1}^B\log P(\sigma^b )
	\end{equation}
	Minimizing $l$ with respect to local fields $h_i$ gives
	\begin{equation}
		\begin{split}
			\frac{\partial l}{\partial h_i(\sigma)} &=\frac{1}{B}\sum_{b=1}^B\left(\frac{\partial\log Z}{\partial h_i(\sigma)}- \delta (\sigma_i^b , \sigma )\right)=
			\\ &=\frac{1}{B}\sum_{b=1}^B\left(f_i (\sigma ) - \delta (\sigma_i^b , \sigma)\right)
			\\&= f_i (\sigma)-F_{i}(\sigma )=0
		\end{split}
	\end{equation}
	Similarly, minimizing $l$ with respect to the couplings gives
	\begin{equation}
		\begin{split}
			\frac{\partial l}{\partial J_{ij}(\sigma,\tau)} &= f_{ij} (\sigma,\tau )-F_{ij}(\sigma,\tau )=0
		\end{split}
	\end{equation}
These equations show that the model with the maximum likelihood to reproduce the sequences observed in the MSA
is the one with frequencies identical to those observed in the MSA.

	A possible strategy to minimize $l$ is \textit{gradient descent}, that is an iterative algorithm
	in which parameters are adjusted by forcing them to follow the opposite direction of the function gradient \cite{ackley1987learning,sutto2015residue,barrat2016improving,haldane2016structural,figliuzzi2018pairwise}.
	The value of the parameters {\boldmath$\theta$} at iteration $k + 1$ can be obtained from the value of
	{\boldmath$\theta$} at the iteration $k$ as
	\begin{equation}
		\theta_{t+1}=\theta_t-\eta_t \nabla_{\theta}l(\theta)=\theta_t-\eta_t (f(\theta)-F)
	\end{equation}
	where $\eta_t$ is the \textit{learning rate} and $t$ is the fictitious time, corresponding to the iteration number.
	Calculation of the gradient requires evaluation of an average over all the possible sequences. This average can be computed with a Metropolis-Hastings algorithm in sequence space,
	but might be very expensive due to the large size of the sequence space.
	In addition, the average should be recomputed at every iteration. 
We here propose to use the instantaneous value of $\delta(\sigma_i,\sigma)$, where
$\sigma_i$ is the identity of the nucleotide at position $i$ in the simulated sequence, as an unbiased estimator of
	$f_i(\sigma)$ in order to update the parameters more frequently, resulting in a \emph{stochastic gradient descent} procedure that
	forces the system to sample the posterior distribution. The procedure can be easily parallelized, so that at each iteration the new set {\boldmath$\theta$} is an average of the updated parameters over all processes.
We here used 20 simultaneous simulations initialized from 20 random
sequences chosen in the MSA.
Once parameters are stably fluctuating around a given value, their optimal value can be estimated by taking a time average of {\boldmath$\theta$} over a suitable time window \cite{cesari2018using}. At that point, a new simulation could be performed using
the time-averaged parameters.
	Such a simulation can be used to rigorously validate the obtained parameters.

	We here choose a learning rate $\eta_t$ in the class \textit{search then converge} \cite{darken1992towards}:
	\begin{equation}
		\label{learn}
		\eta_t = \frac{\alpha}{1+\frac{t}{\tau_S}}
	\end{equation}
	This function is close to $\alpha$ for small $t$ (``search phase''). For $t\gg\tau_S$ the function decreases as $1/t$ (``converge phase'').
	Since it is based on Boltzmann sampling of the sequence space, we refer to this procedure as \textit{Boltzmann learning}.
	The exact algorithm is described in Supporting Information and the employed C code is available at https://github.com/bussilab/bl-dca.
We notice that in the algorithm implemented here, at variance with others proposed before \cite{sutto2015residue,figliuzzi2018pairwise}, the
Lagrangian multipliers are evolved every few Monte Carlo iterations using istantaneous values rather than
averages obtained from converged trajectories. In Ref.~\cite{figliuzzi2018pairwise} a
change of variables of the model parameters was proposed to make the minimization easier. This idea might be beneficial also
in our algorithm.

\subsection{Capability of the inferred couplings to reproduce frequencies}
The capability of various DCA methods to infer correct parameters for the Potts model can be quantified
by computing the root-mean-square deviation (RMSD)
       between model and observed pair frequencies:
       \begin{equation}
               RMSD=\sqrt{<(f_{ij} (\sigma_i , \tau_j )-F_{ij} (\sigma_i , \tau_j ))^2>_{\{ij\},\{\sigma_i,\tau_j\}}}
       \end{equation}
For Boltzmann-learning DCA, the model frequencies are calculated in the validation phase of simulations, and the RMSD can be used
to assess the convergence of the simulation.
For other DCA methods one can simply use the estimated couplings to run a simulation in sequence space.
	\subsection{Validation of the predicted contacts}
	We perform this analysis on sequences of a number of riboswitches families classified in the Rfam database \cite{nawrocki2014rfam}.
Columns with more than 90\% of gaps were removed from the alignments
	in order to make the maximization faster and to avoid overfitting the model on positions of the alignment that are not relevant.
		The 17 RNA families have been chosen among those for which at least one
		high-resolution crystallographic structure have been reported, ruling out from the analysis the structures annotated as interacting double chains. A full list is reported Supporting Information (Table 1).
The number of nucleotides in each chain ranges between 52 and 161, and the effective number of sequences
between 25 and 1078 (all details are reported in Supporting Information, Table 1).
The lowest quality structure in the data set has been solved with resolution 2.95\AA.
Contacts in the reference PDB structures are annotated with DSSR \cite{gkv716},
that takes into account all hydrogen bonds and classify base pairs according to the
Westhof-Leontis nomenclature \cite{leontis2001geometric}.
This is different from other works where the geometric distance between heavy atoms belonging to each nucleotide, thus including
also backbone atoms,
is used, and is expected to better report on the direct base-base contacts that are supposed to be associated to covariation.
We decided to ignore stacking interactions since
coevolution in RNA is mostly related to isostericity \cite{leontis2002non,stombaugh2009frequency}.
All the used MSAs as well as files containing the annotation of each
base pair are available at https://github.com/bussilab/bl-dca.

Before computing the one-site and two-sites frequencies,
the columns of the MSA where the sequence corresponding to the
reference crystallographic structure had a gap were eliminated by the
alignment. Whereas this step should not be in principle required,
preliminary calculations showed that this pruning improves the
quality of the results for all the tested DCA methods
(data not shown).
In addition, we applied to the score of each contact (Eq.~SI5) the so-called average-product correction
(APC) \cite{dunn2007mutual}.

	Evaluation of the performance of RNA contact prediction methods requires the number of correct predictions (true positives, TP), the number of contacts predicted but absent in the native structure (false positives, FP), and the number of contacts present in the native structure but not predicted (false negatives, FN). Two common measures are sensitivity and precision, where $\textit{sensitivity}$ is the fraction of correctly predicted base pairs of all true base pairs, while $\textit{precision}$ is the fraction of true base pairs of all predicted base pairs:
	\begin{align}
		sensitivity&=\frac{TP}{TP+FN}\\
		precision&=\frac{TP}{TP+FP}
	\end{align}
	The Matthews correlation coefficient (MCC) can be defined as the geometric average of sensitivity and precision \cite{matthews1975comparison,Gorodkin2001DiscoveringCS}
	\begin{equation}
		MCC=\sqrt{sensitivity\cdot precision}
	\end{equation}
	and is equivalent to the interaction network fidelity \cite{parisien2009new}.
	To turn contact scores $S_{ij}$ (either Eq.~SI7, for mutual information, or Eq.~SI5, for DCA, or E-values, for R-scape) into predictions it is necessary to assume a threshold $\bar{S}$.
	The predicted contacts will be those scored by a value above (below, for R-scape) $\bar{S}$.
	For R-scape, we used the recommended threshold $\bar{S}=0.05$.
	For the other methods, we chose the threshold score maximizing the MCC, corresponding to the optimal compromise between precision and sensitivity.
	For each covariance method, the MCC as a function of the threshold score $S$ shows a similar behavior for all the $N_s$=17 systems, their peaks falling at very similar positions. This suggests the possibility to set a unique threshold for each covariance method that maximizes the MCC geometric average over all systems:
	\begin{equation}
		\label{thresh}
		\bar{S}=\arg\max_{S} \left(\prod_{\mu}^{N_s}MCC_{\mu}(S)\right)^{\frac{1}{N_s}}
	\end{equation}
	\section{Results}
We here report an extensive assessment of the capability of covariance-based methods to infer contacts in RNA systems.
In particular, we focus on direct-coupling-analysis (DCA) methods, which require the coupling constants of
a Potts model that reproduces empirical covariations to be estimated. We thus first assess the capability of
different methods to infer correct couplings. We then compare the high-score contacts with those observed in high resolution
crystallographic structures in order to assess the capability of these methods to enhance RNA structure prediction.

The majority of the results presented in the main text are obtained using the Infernal MSA,
and equivalent results obtained using ClustalW alignments are presented in Supporting Information (Figures 6 to 11).
Similarly, the effect of not applying the average-product correction (APC) is reported in Supporting Information (Table 4).

\subsection{Capability of the inferred couplings to reproduce frequencies}

\begin{figure}
				
			\begin{minipage}[b]{4.2cm}
	\includegraphics[width=\textwidth]{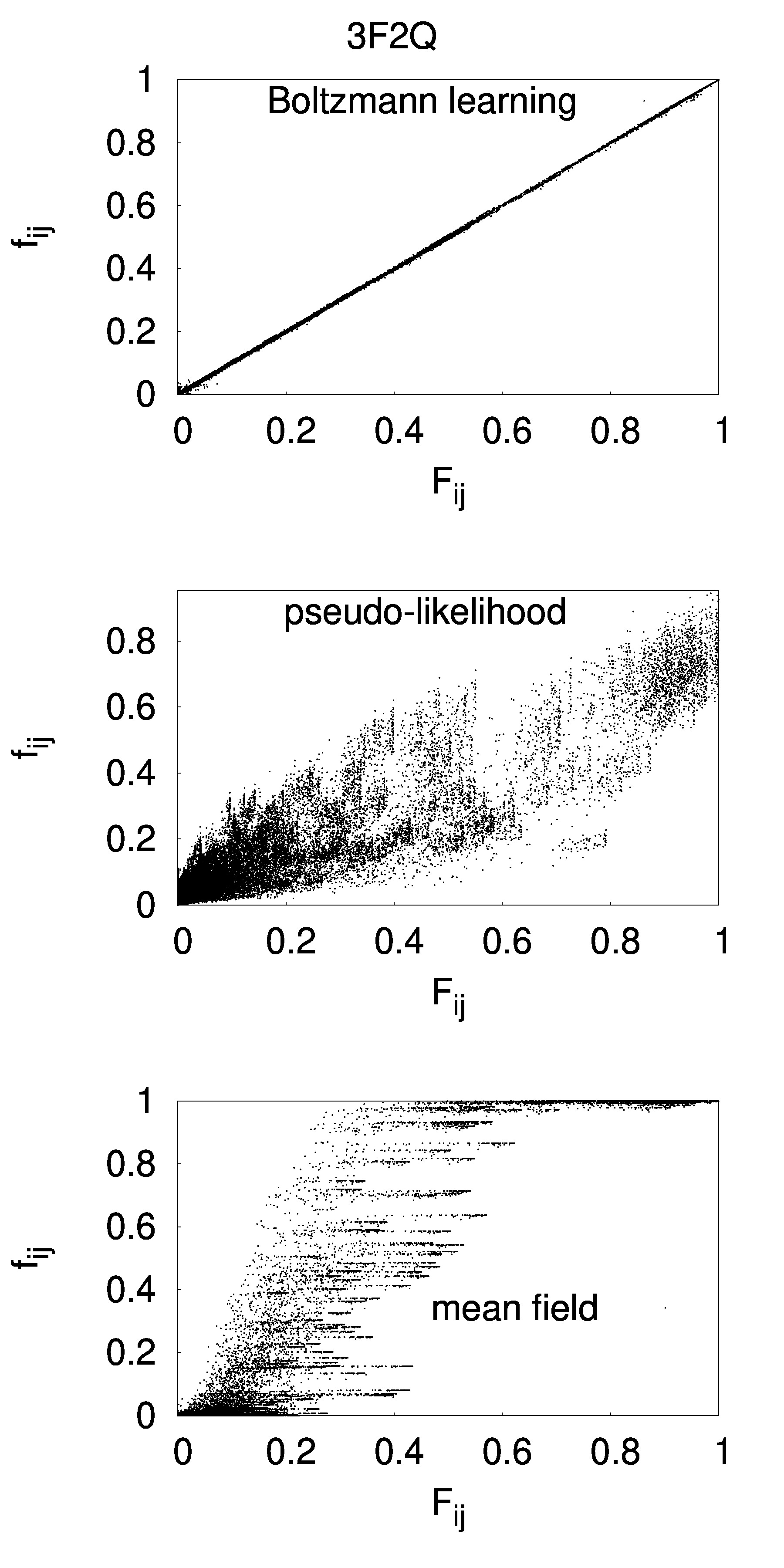}
			\end{minipage}
					\begin{minipage}[b]{4.2cm}
				\includegraphics[width=\textwidth]{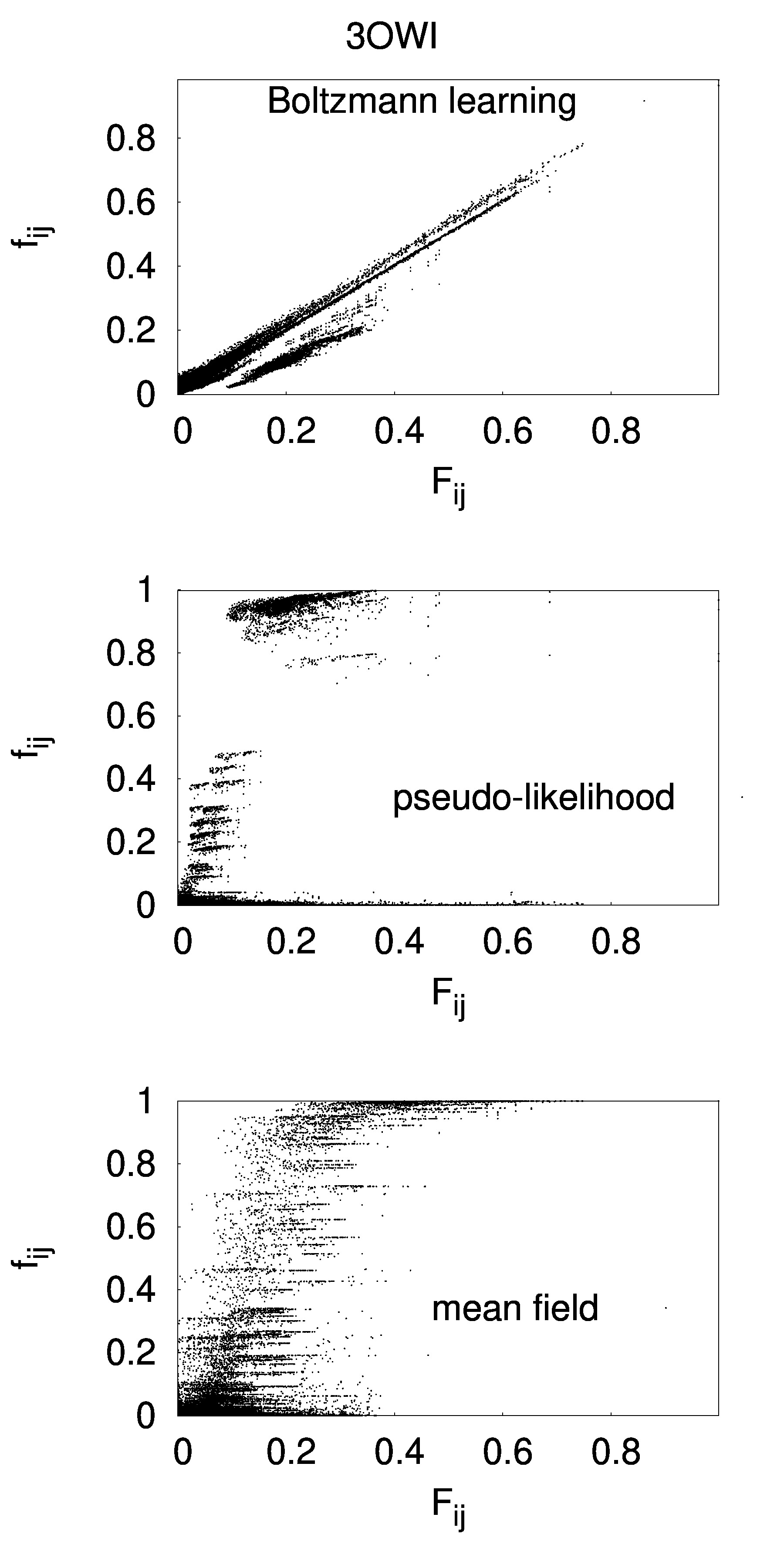}
			\end{minipage}	
						\caption{FMN riboswitch (PDB code 3F2Q) (\ref{fij}) and c-di-GMP-I (PDB code 3IRW) (\ref{mffij}). Comparison between  modeled $f_{ij}(\sigma,\tau)$ and empirical $F_{ij}(\sigma,\tau)$ frequencies $\forall i,j,\sigma,\tau$, obtained from DCA via Boltzmann learning, mean-field approximation and pseudo-likelihood maximization.}
			\label{fij}
				\label{mffij}
		\end{figure}

As a first step, we compared the absolute capability of the discussed methods to infer a Potts model
compatible with the frequencies observed in the MSA.
As shown in Figure \ref{fij}, the Boltzmann learning procedure is capable to infer a Potts model that generates sequences
with the correct frequencies. The two displayed families are those where the model frequencies agree best (PDB: 3F2Q) or worst (PDB: 3IRW)
with the empirical ones. For 3IRW there are still visible mismatches, whereas for 3F2Q the modeled and empirical frequencies are virtually identical.
On the other hand, the couplings inferred using the pseudo-likelihood or the mean-field approximation do not reproduce correctly the empirical frequencies.
This is expected, since the mean-field approximation is not meant to be precise but rather a quick method to compute an approximation to
the real couplings.
Particularly striking is the case of the pseudo-likelihood for 3IRW, where there is no apparent correlation between the modeled and the empirical frequencies.

\begin{figure}
		\includegraphics[width=\linewidth]{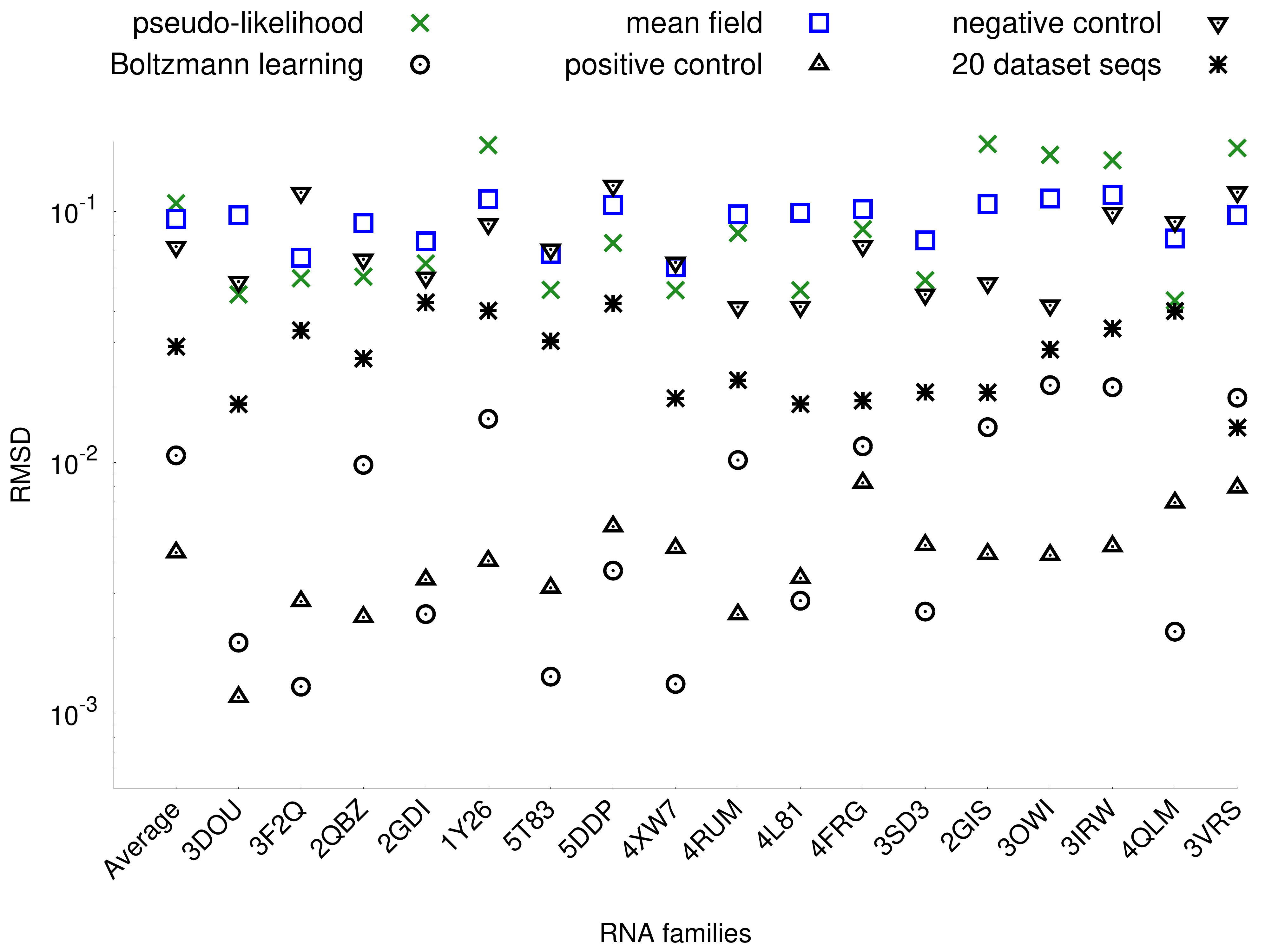}
		\caption{
Capability of the inferred couplings to reproduce frequencies using different methods (Boltzmann learning, pseudo-likelihood and mean-field DCA).
The validation is done running a parallel MC simulation
on 20 sequences and calculating the root-mean-square deviation (RMSD) between the obtained frequencies and the empirical ones.
We report a positive control (statistical error due to the finite number of sequence),
a negative control (RMSD between empirical sequences and a random sequence) and
the RMSD from the ensemble of the 20 sequences used as a starting point of Boltzmann learning simulations.
Families are labeled using the PDB code of the
representative crystallographic structure. Average RMSD is also reported.}
		\label{conv}
	\end{figure}

In Figure \ref{conv} we report the RMSD between the empirical and model frequencies for all the investigated families. The learning parameters for the Boltzmann learning simulation were chosen in order to minimize the RMSD value reported here ($\alpha=0.01$, $\tau_S=1000$).
 A negative control is performed comparing empirical frequencies with the ones calculated on random sequences
($f_{ij}=1/25$), and a positive control computing the statistical error due to finite size of the alignment, in order to set a reference for RMSD values. In addition, we compare empirical frequencies against the ones calculated on the 20 MSA sequences initializing the parallelized Boltzmann learning simulation, so to ensure that frequencies are not reproduced thanks to the statistics resulting from the initial sequences
but rather thanks to a correct choice of the coupling parameters.
For all families, the resulting RMSD obtained with the Boltzmann learning couplings is lower than the one obtained using the 20 sequences from the MSA,
indicating that the chosen couplings are shifting the distribution towards the empirical one. In some cases the RMSD reaches
the statistical error expected with a finite number of sequences (positive control).
Whereas this is expected since the Boltzmann learning procedure is exaclty trained to reproduce these frequencies, it is not obvious that this result can be achieved
in a feasible computational time scale.
On the contrary, both the pseudo-likelihood and mean-field approximation present an RMSD systematically larger than the one obtained from 20 sequences from the MSA.
This indicates that the couplings inferred using these approximated methods are not leading to a Potts model that reproduces
the experimental frequencies.

We notice that the adopted pseudo-likelihood implementation employs a regularization term in order
to improve predictions when the number of sequences is low.
This term is usually tuned in order to improve the rank of true contacts and not the frequencies reported here.
We thus tested parameters
obtained using a lower regularization term obtaining similar results (Supporting Information, Figure 12).
Given that pseudo-likelihood is known to converge to the exact value in the limit of an infinite number of sequences
(see, e.g., \cite{arnold1991pseudolikelihood} and \cite{ravikumar2010high}),
this discrepancy should be attributed to the typical size of the used alignments.
We also notice that in multiple cases the frequencies obtained using couplings inferred with pseudo-likelihood tend to be larger than the empirical ones. Since the RMSD is highly sensitive to large deviations, this can cause some of the systems to be in less agreement with natural sequences than the employed negative control, which instead consists by construction of homogeneous frequencies.
Qualitatively, the deviation observed here is similar to the one reported for protein systems
in \cite{figliuzzi2018pairwise}.


\subsection{Validation of contact prediction}

As we have seen so far, Boltzmann learning is the only procedure capable to infer correct couplings.
However, this does not necessarily imply that it is also the method capable of most correct contact predictions.
Indeed, one cannot give for granted that the exact parameters of the Potts model are correlated with structural contacts.
We here validate the predictions against a set of crystallographic structures by computing the MCC between the predicted and empirical contacts.
The general approach used to predict contacts from DCA is to extract the residue pairs with the highest couplings.
Similarly, contacts can be predicted choosing pairs with the highest mutual information or the lowest E-value provided by R-scape.
In order to fairly choose the threshold we adopted a cross-validation procedure: the $\overline{MCC}$ of each system is the one corresponding to a score cutoff $\overline{S}$ maximizing the average MCC (Eq.~\ref{thresh}), calculated excluding that system. The choice of the threshold for covariance scores of the different models can be generalized to an independent data set, since the optimal threshold has a similar value for all systems (Supporting Information, Table 2 and 3).  
We also tested the more standard procedure of choosing as predictions a given fraction of the length $N$
(Supporting Information, Table 11).
For R-scape we used the recommended threshold corresponding to an E-value equal to 0.05.

	As a  negative control we show the MCC obtained assuming randomly chosen scores. In this case,
the precision is equal to the number of native contacts ($N_{native}$) over the total number of possible contacts ($\frac{N(N-1)}{2}$) irrespectively of
the chosen threshold,
whereas the sensitivity is maximized when the threshold is chosen such that all the possible contacts are predicted and is equal to 1.
The corresponding $MCC$ is thus $\sqrt{\frac{2N_{native}}{N(N-1)}}$.
Finally, we also computed the cross-validated MCC obtained with a thermodynamic model applied to the sequence
associated to each crystallographic structure, by using as scoring the pairing probabilities
computed with ViennaRNA \cite{mathews2004incorporating,lorenz2011viennarna}. These results do not exploit the covariance information
and  are thus instructive to assess its importance.

\begin{figure}[h]
		\includegraphics[width=\linewidth]{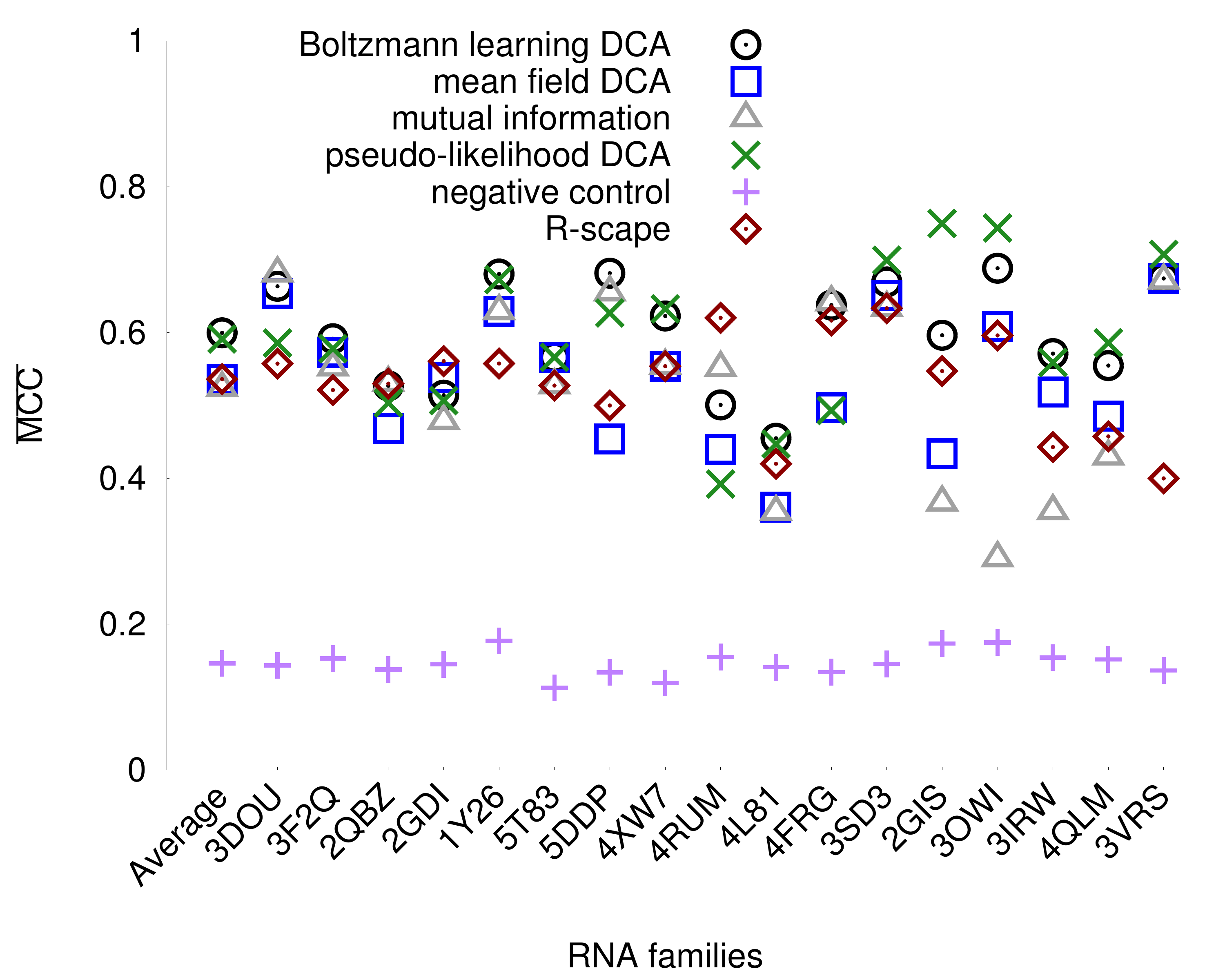}
		\caption{$\overline{MCC}$ of Boltzmann learning DCA, pseudo-likelihood DCA, mean-field DCA, mutual information, and R-scape for 17 RNA families at the threshold obtained through cross-validation procedure. The recommended threshold 0.05 was used for R-scape. Families are labeled using the PDB code of the representative crystallographic structure. Average $\overline{MCC}$ is also reported. Alignments are performed with $\textit{Infernal}$.
}
		\label{mcc}
	\end{figure}

	Results of the cross-validation procedure for each system (Figure \ref{mcc}) indicate that direct coupling analysis outperforms mutual information and R-scape, and in particular Boltzmann learning performs the most accurate prediction.
In addition, the results on individual families show that the choice of threshold covariance score is more consistent for Boltzmann learning when compared to pseudo-likelihood DCA. In order to quantify this effect we introduce a transferability index $\phi=\frac{1}{N_s}\sum\limits_{\mu}^{N_s}\frac{\overline{MCC_{\mu}}}{MCC_{\mu}^{max}}$,which is the ratio between the
cross-validated MCC for system $\mu$ ($\overline{MCC}_{\mu}$)
described above and the maximum MCC that can be obtained by choosing the optimal threshold for each system $MCC_{\mu}^{max}$,
averaged over all systems.
This value amounts to $\phi=0.96$ for BL and to $\phi=0.91$ for pseudo-likelihood DCA,
suggesting that for the latter case the accuracy of contact prediction is more sensible to the choice of the cutoff, which is less easily transferable between different systems.
Results for mean field DCA and mutual information are $\phi=0.95$ and $\phi=0.92$, respectively.
Finally, we notice that all the DCA methods perform better than thermodynamic models alone (Supporting Information, Table 8).
	\begin{figure}
		\includegraphics[width=0.9\linewidth]{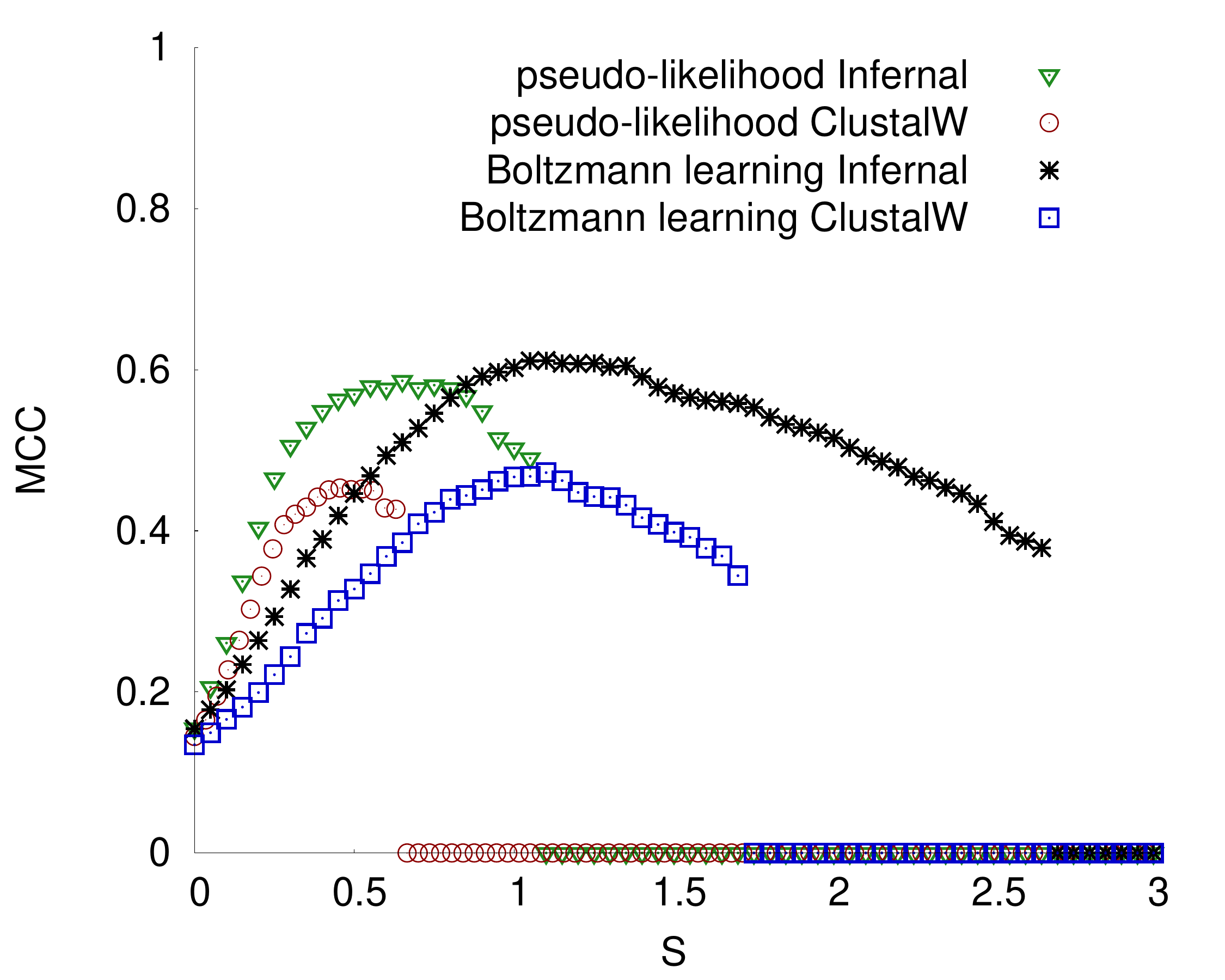}
		\caption{Geometric-average MCC as a function of threshold scores S for Boltzmann learning and pseudo-likelihood DCA. MSAs are performed with $\textit{ClustalW}$ and $\textit{Infernal}$, as indicated. The sharp decrease after some method-dependent value of S is due to the fact that when the threshold is
			too large the number of correctly predicted contacts in at least one of the 17 investigated systems drops to zero.}
		\label{al}
	\end{figure}
	\subsection{Influence of alignment method}
We then used the two most accurate covariance methods (Boltzmann learning and pseudo-likelihood DCA)
to assess the influence of the alignment method. In particular, we considered the MSA methods 
 implemented in $\textit{ClustalW}$ and $\textit{Infernal}$ packages.  The average MCC over all RNA families when varying threshold S is systematically higher if sequences are aligned with $\textit{Infernal}$ rather than $\textit{ClustalW}$ (Figure \ref{al}). We attribute this improvement in the quality of prediction performance to the use of consensus secondary structure in $\textit{Infernal}$ \cite{nawrocki2013infernal}.
The discrepancy between the accuracies of contact prediction using two different alignment methods enlightens the necessity of efficient tools to improve covariance analysis input quality.
	Interestingly, the threshold score $\overline{S}$ maximizing the MCC is the same for the Boltzmann learning performed on the two different MSAs. This suggests the robustness of the adopted procedure to assess the optimal threshold score (Eq.~\ref{thresh}), again enlightening a greater consistency in its choice for the Boltzmann learning  with respect to pseudo-likelihood maximization framework.
	Given its better performance, the $\textit{Infernal}$ MSA method is used in the rest of the main text.

	\subsection{Precision and sensitivity}
	In order to better quantify the capability of the investigated methods to provide useful information about contacts, we independently monitor
	sensitivity and precision for each RNA family at cross-validation thresholds.
	The average sensitivity values are around 0.3--0.4, indicating that approximately one third of the contacts present in the native
	structure can be predicted with these procedures (Supporting Information, Figure 1).
Qualitatively, it appears that correctly predicted contacts are scattered along the sequences.
The average precision instead ranges between 0.7 and 0.9, indicating that the number
	of falsely predicted contacts is rather small (Supporting Information, Figure 2). The Boltzmann learning and pseudo-likelihood DCA report higher sensitivity and precision
	than the other methods.
R-scape presents a higher sensitivity when compared with mutual information and a similar precision.
We notice that R-scape results reported here are obtained using the recommended threshold (E-value$<$0.05).
Results obtained choosing the E-value that maximizes the MCC are reported in Supporting Information (Tables 9 and 10).
	In order to assess the capability of these methods to probe RNA tertiary structure we also report
	the sensitivity value restricted to secondary contacts, obtained considering only base pairs contained in stems,
and the number of true positive tertiary contacts, with results similar to those reported above (Supporting Information, Figures 3 and 4).
A contact is thus here considered as tertiary irrespectively of which edges are shared between nucleobases, and might even
be an isolated WC pair.
In general, DCA is able to identify not only cWW \cite{leontis2001geometric} pairs, where covariance is mostly associated to
canonical pairs (GC, AU, and GU), but also a number of non-canonical pairs (see Table 6).
When looking at the absolute number of incorrect predictions the Boltzmann learning DCA provides the smallest average number (Supporting Information, Figure 5).
In particular, pseudo-likelihood DCA reports a very large number of false positives for a few systems.
Also in this case, this is a consequence of the poor transferability of the cutoff for contact prediction in pseudo-likelihood DCA.
	A more careful eye on incorrect predictions reveals that $\approx 50\%$ of false positives predicted by all DCA methods are actually stacking interactions not included in the true-positive list since we only considered base pairings in reference native structures (Supporting Information, Table 7).
In addition, couplings in consecutive nucleotides might be affected by a bias in the dinucleotide distribution.

\subsection{Typical contact predictions}

\begin{figure}
				
				\begin{minipage}[b]{4.2cm}
					\includegraphics[width=\textwidth]{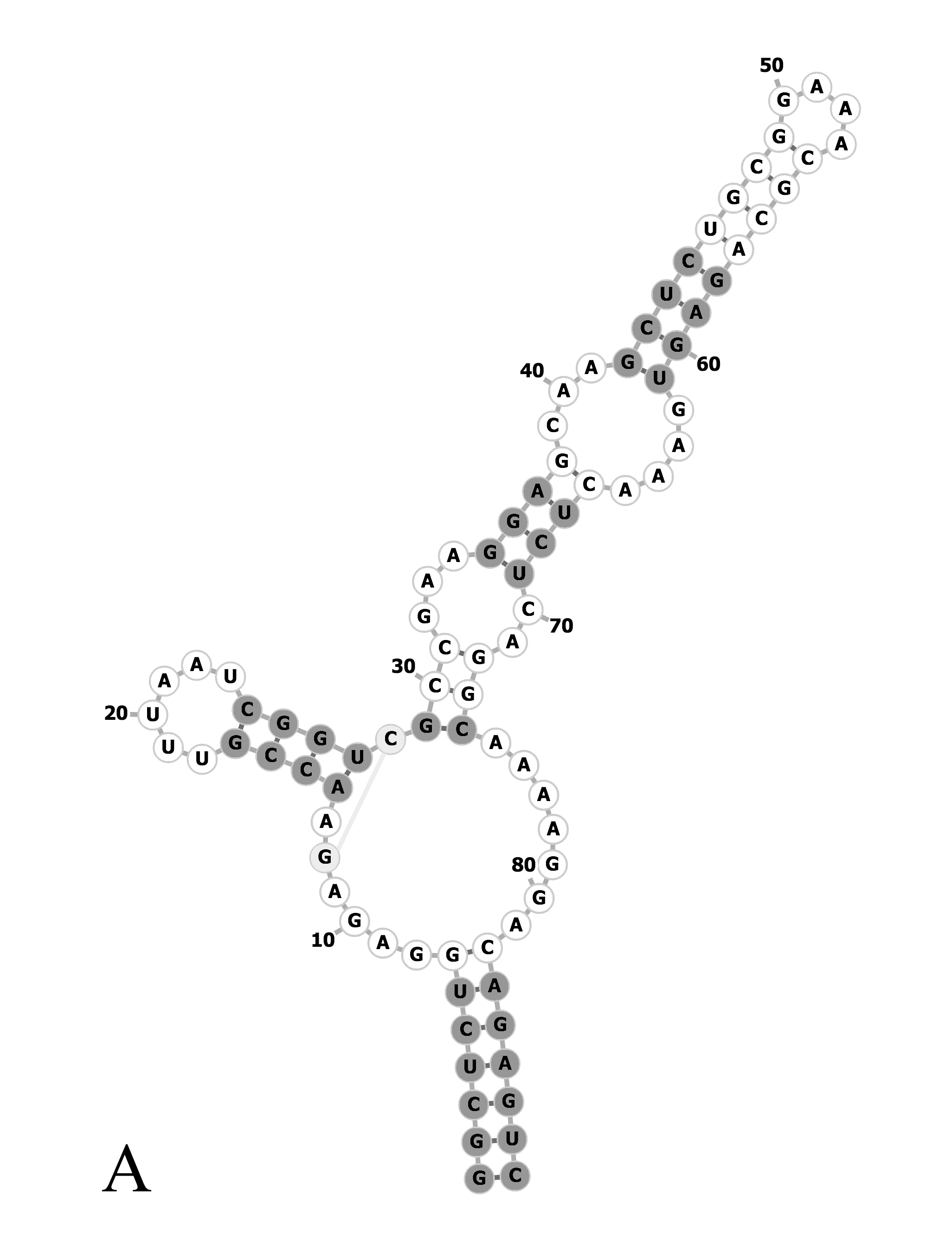}
					\label{a}
				\end{minipage}
				\begin{minipage}[b]{4.2cm}
					\includegraphics[width=\textwidth]{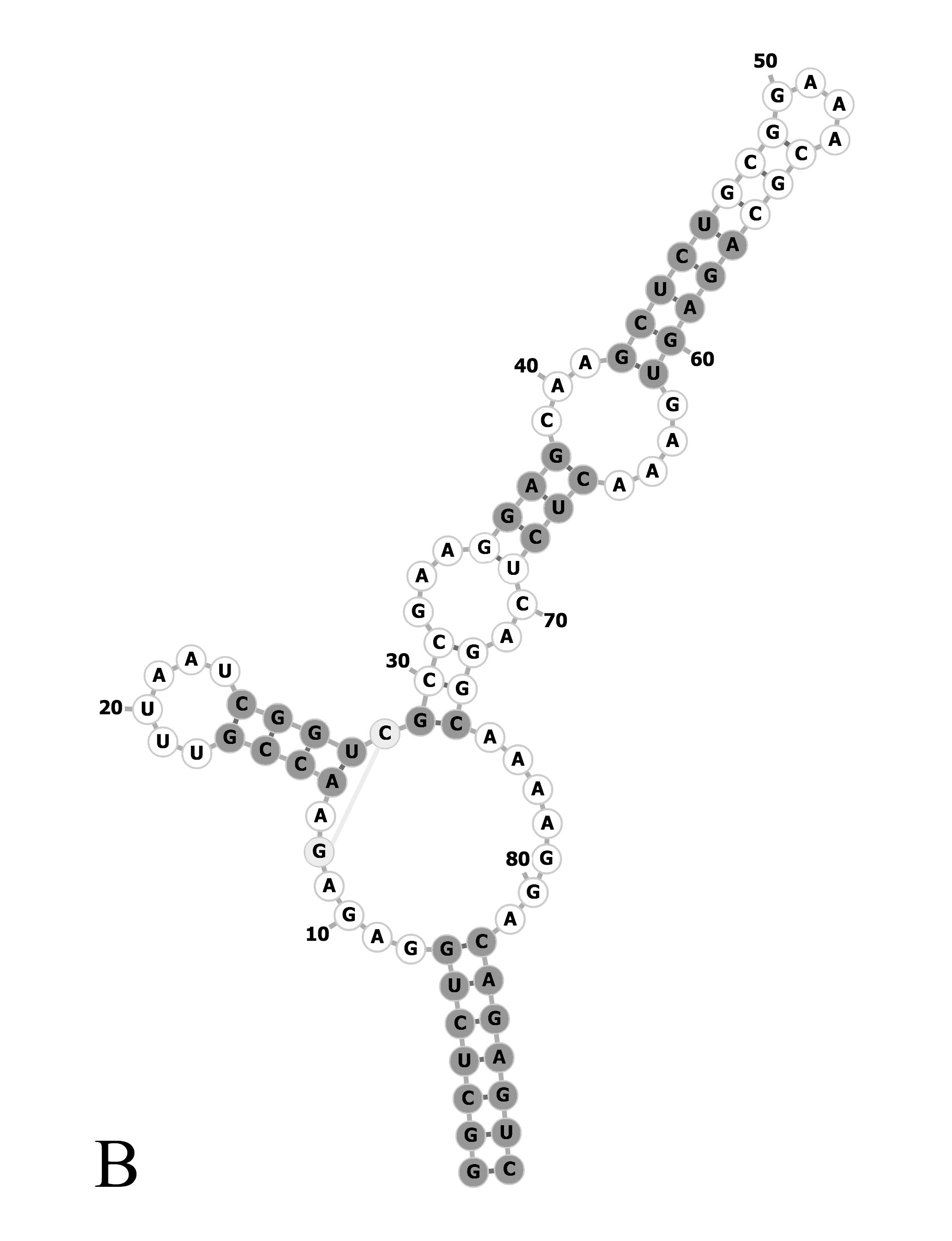}
					\label{b}
				\end{minipage}

				\caption{Glycine riboswitch (PDB code 3OWI) most accurate Boltzmann learning prediction (5A) and respective pseudo-likelihood prediction (5B). Correctly predicted contacts in secondary structure are
					shown in dark-grey. Correctly predicted tertiary contacts are shown in light-grey.
We notice that G12/C28 pair is here labeled as tertiary since it corresponds to a isolated Watson-Crick pair in the reference structure.}
							\label{gly}		
			\end{figure}

\begin{figure}
			
			\begin{minipage}[b]{4.2cm}
				\includegraphics[width=\textwidth]{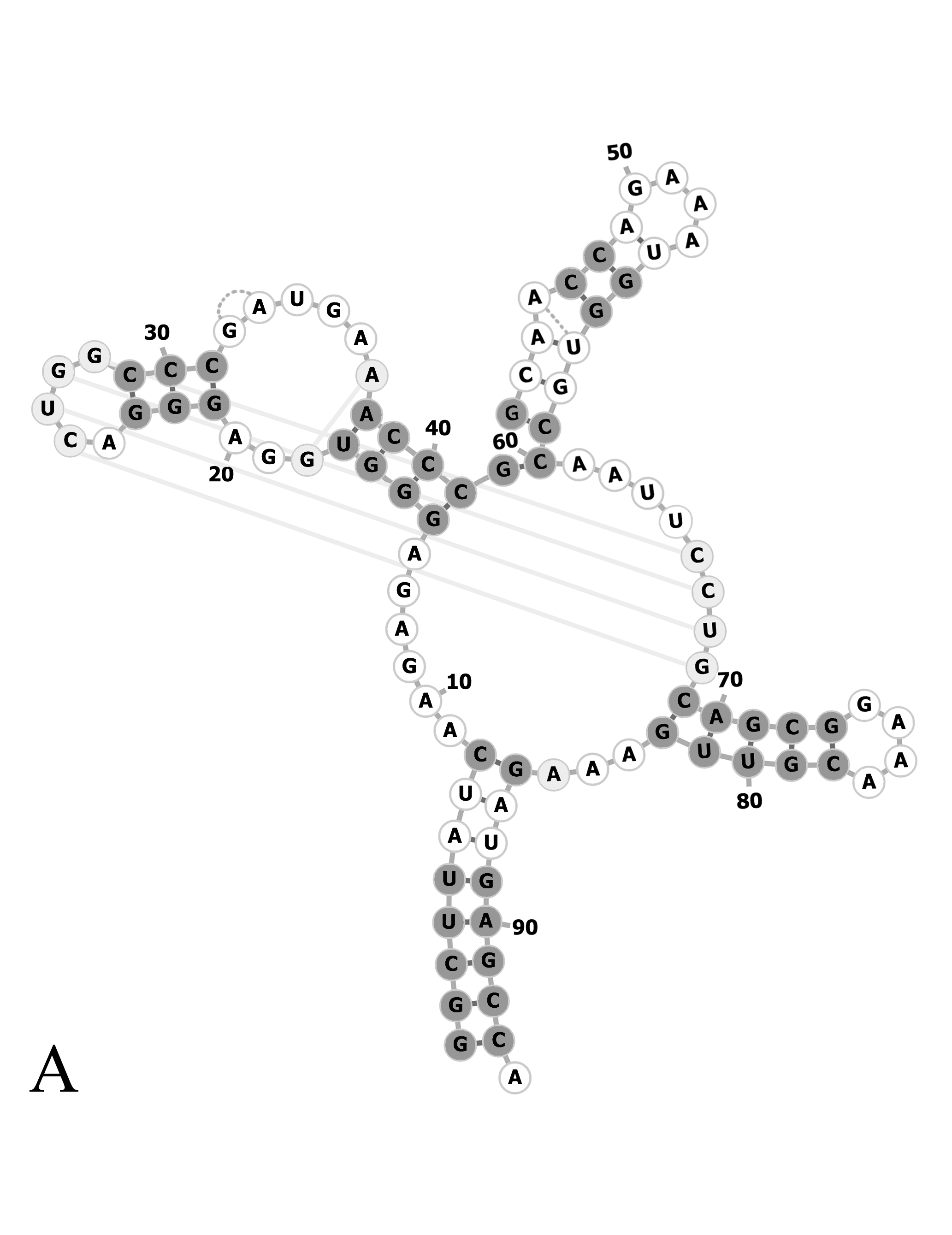}
				\label{e}
			\end{minipage}
			\begin{minipage}[b]{4.2cm}
				\includegraphics[width=\textwidth]{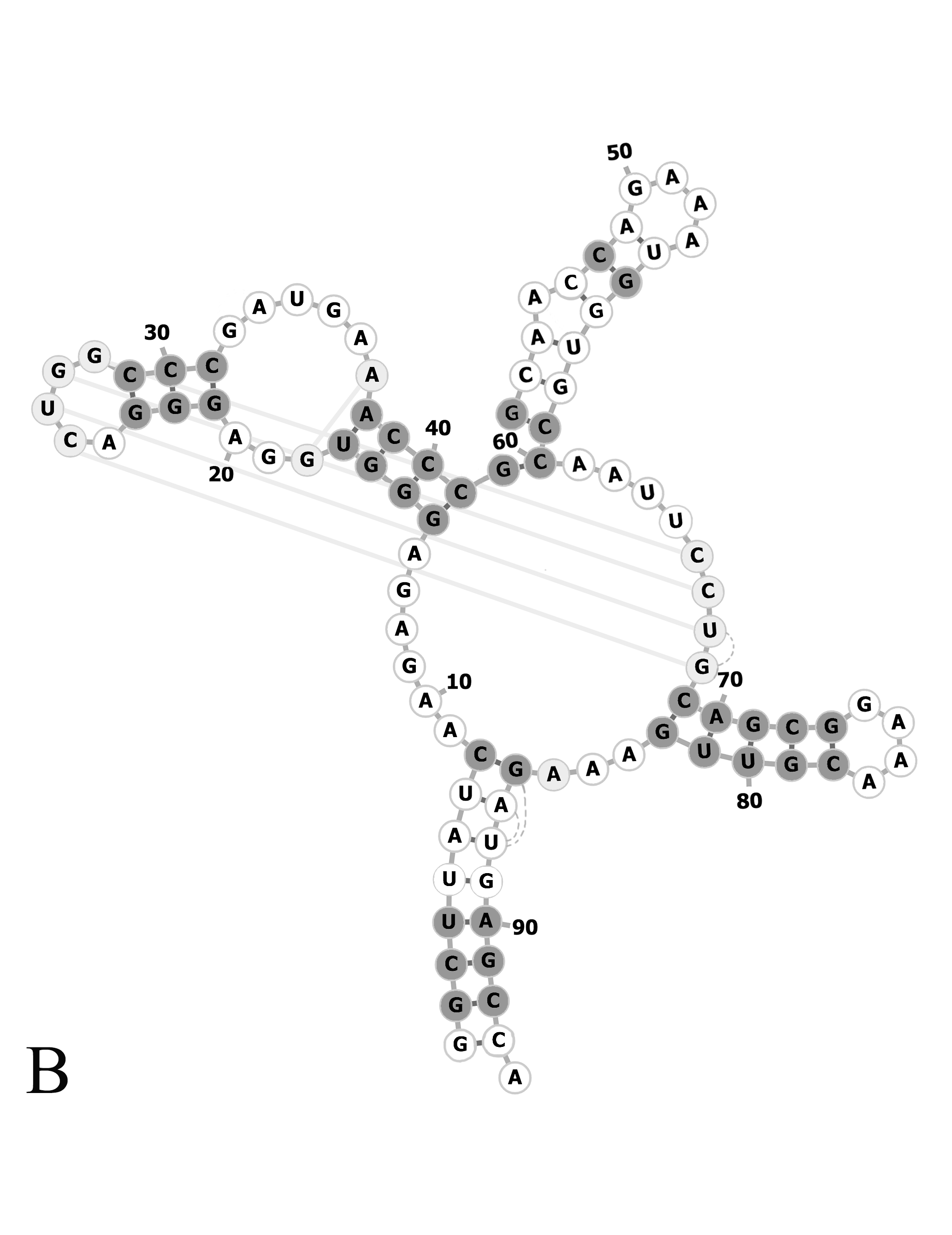}
				\label{f}
			\end{minipage}	
			\caption{	SAM riboswitch (PDB code 2GIS), best accurate pseudo-likelihood prediction (6A) and respective Boltzmann learning prediction (6B).
				Correctly predicted contacts in secondary structure are
				shown in dark-grey. Correctly predicted tertiary contacts are shown in light-grey. False positives are shown with dashed lines.}
			\label{samm}
		\end{figure}

\begin{figure}
					\begin{minipage}[b]{4.0cm}
						\includegraphics[width=\textwidth]{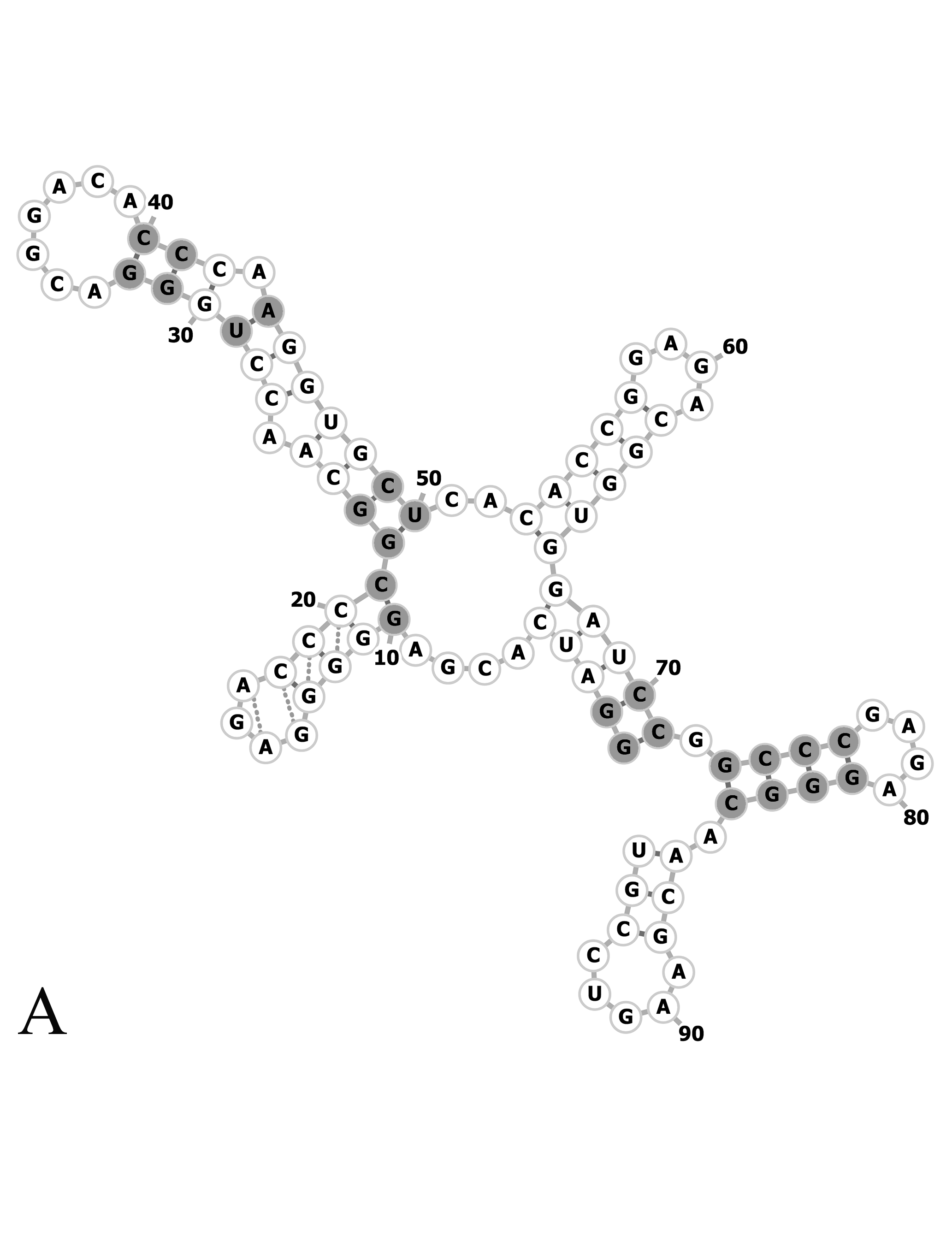}
						\label{c}
					\end{minipage}
					\begin{minipage}[b]{4.0cm}
						\includegraphics[width=\textwidth]{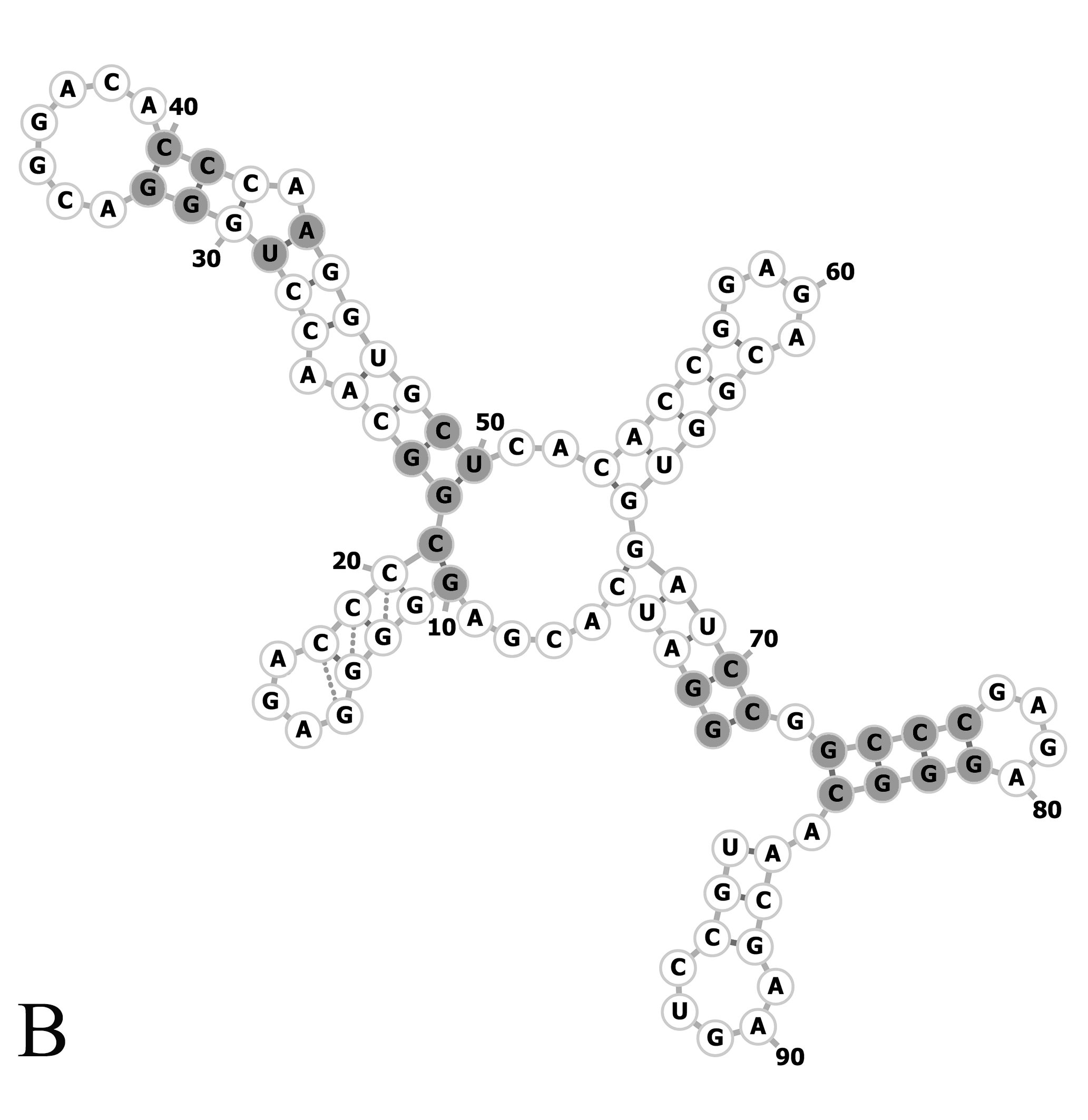}
						\label{d}
					\end{minipage}	
					\caption{	SAM-I/IV riboswitch (PDB code 4L81), least accurate Boltzmann learning prediction (7A) and respective pseudo-likelihood prediction (7B).
						Correctly predicted contacts in secondary structure are
						shown in dark-grey. Correctly predicted tertiary contacts are shown in light-grey. False positives are shown with dashed lines.}
					\label{sam}
				\end{figure}

\begin{figure}
			
			\begin{minipage}[b]{4.2cm}
				\includegraphics[width=\textwidth]{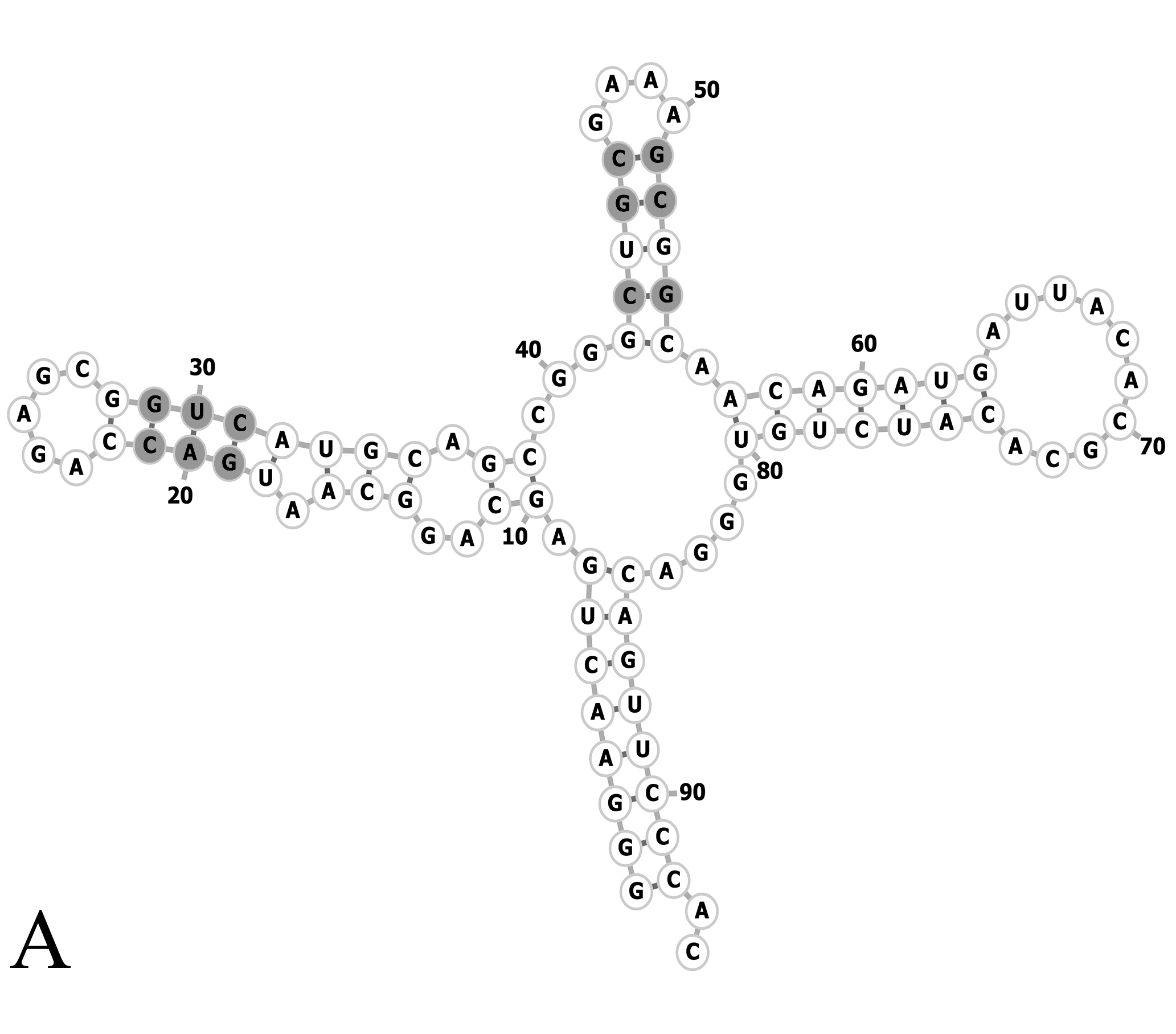}
				\label{g}
			\end{minipage}
			\begin{minipage}[b]{4.2cm}
				\includegraphics[width=\textwidth]{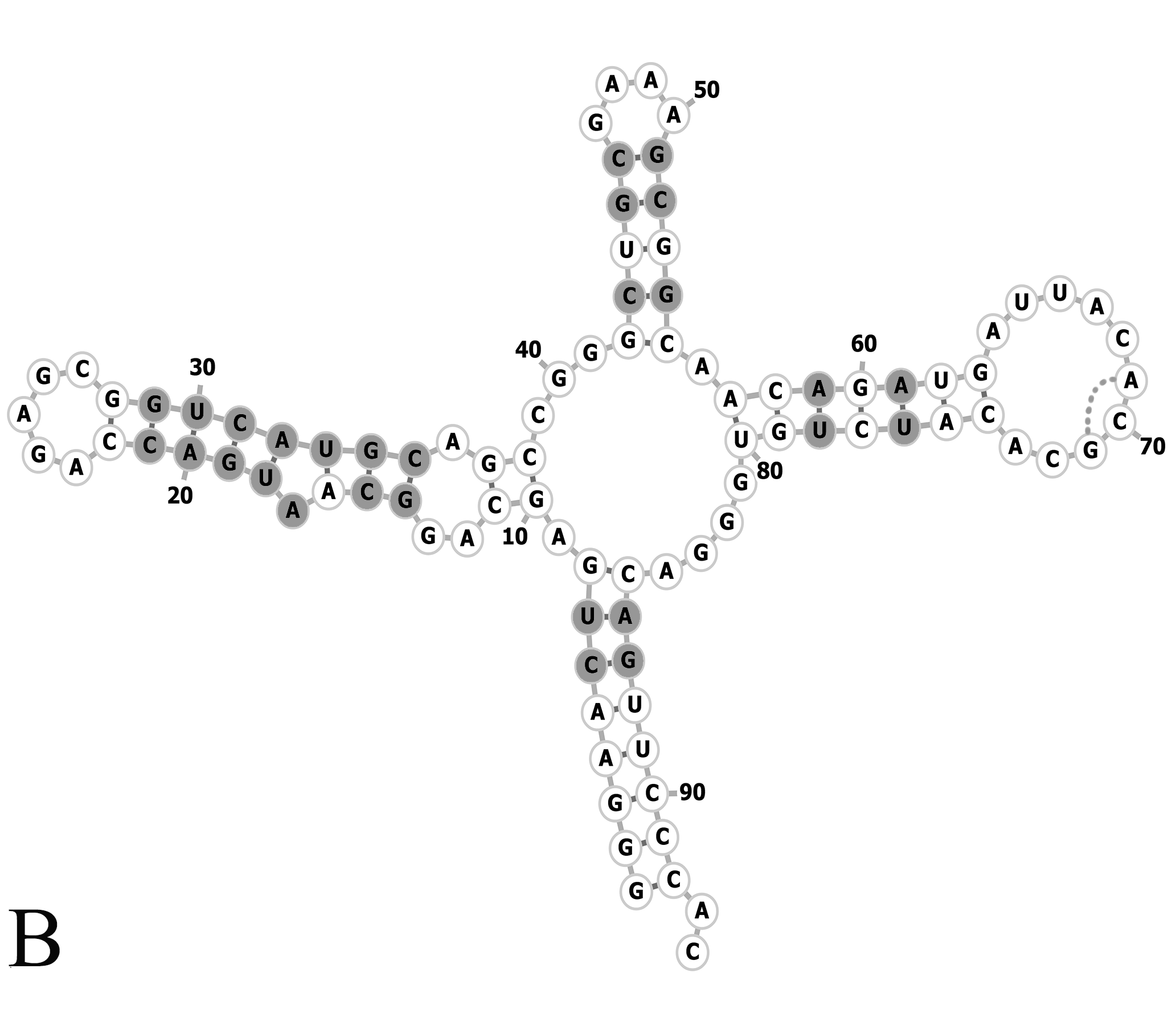}
				\label{h}
			\end{minipage}	
			\caption{		NiCo riboswitch (PDB code 4RUM), least accurate pseudo-likelihood prediction (8A) and respective Boltzmann learning prediction (8B).
				Correctly predicted contacts in secondary structure are
				shown in dark-grey. Correctly predicted tertiary contacts are shown in light-grey. False positives are shown with dashed lines.}
			\label{nic}
		\end{figure}

It is instructive to visualize which specific contacts are correctly predicted and which ones are not for individual systems.
We first discuss the predictions on the systems where Boltzmann learning and pseudo-likelihood DCA 
result in the highest MCC (glycine riboswitch, PDB 3OWI, and SAM riboswitch, PDB 2GIS, respectively).
In the glycine riboswitch, Figure \ref{gly}, we see that the two methods give comparable results. All the four native stems are predicted, although
pseudo-likelihood DCA predicts a slightly larger number of correct pairs. Also a non-stem WC contact is identified.
In the SAM riboswitch, Figure \ref{samm}, we see that the pseudo-likelihood DCA predicts a significantly larger number of correct contacts.
Notably, both methods are capable to identify contacts in a pseudoknotted helix between residues 25--28 and residues 68--65.
These examples show that in the best cases these methods allow full helices to be identified accompanied by a small number of critical tertiary contacts.
It is also useful to consider the cases resulting in the lowest MCC (SAM-I/IV riboswitch, PDB 4L81, for Boltzmann learning
and NiCo riboswitch, PDB 4RUM, for pseudo-likelihood DCA).
In the SAM-I/IV riboswitch the two methods give comparable results, and 
	only a limited number of secondary contacts are correctly predicted (Figure \ref{sam}).
	The stem between position 10 and position 20 shows a number of false positives. In this case, a helix with
	a register shifted by one nucleotide is suggested by the both DCA predictions.
In more detail, we do not expect the alternative register to have a significant population in solution,
since it would be capped by a AGAC tetraloop, whereas the reference crystal structure displays a common GAGA tetraloop.
We interpret both sets of false positives as errors in the MSA.
Indeed, especially with sequences consisting of consecutive identical nucleotides, one cannot assume the alignment procedure to
correctly place gaps in the MSA. As a consequence, the reference structure for which the PDB is available might be misaligned
with the majority of the homologous sequences in the MSA, resulting in predicted contacts shifted by one position
upstream or downstream.
	Remarkably, many WC pairs close to the binding site of the riboswitch are predicted (G10/C21, G22/U50 and G23/C49;
	ligand directly interacts with nucleotides C7, A25 and U47).
In the  NiCo riboswitch, Figure \ref{nic}, pseudo-likelihood DCA only predicts 6 correct helical contacts, whereas Boltzmann learning DCA is capable to predict a number of contacts in the helices, even though resulting in several
false positives.


\subsection{Validation on non-riboswitch systems}

We further validated the whole procedure by considering 4 additional families including ribosomial RNA subunits, transfer RNA (tRNA), and a purely eukaryotic 
spliceosomal RNA. All the parameters of the Boltzmann learning simulations were chosen identical to those used for the riboswitch families. The threshold
used to convert scores into predictions was taken as 1.06, which is the one that maximizes the MCC on the 17 riboswitch families.
Results are reported in Supporting Information, Table 12 and are slightly worse than those obtained for riboswitch families, with the exception of tRNA.
	
	\section{Discussion}
	We here report a systematic assessment of RNA contact prediction based on aligned homologous sequences using mutual information analysis, R-scape, and DCA.
	When compared to previous works \cite{de2015direct,weinreb20163d,wang2017optimization}, our analysis focuses on the DCA calculation and does not
	convert the resulting couplings into a structural model.
The capability of various DCA-based methods to reproduce empirical frequencies from the MSA is evaluated.
Native contacts in a set of reference structures are carefully annotated and compared with the predicted ones, in order to quantify the fraction
	of correctly predicted contacts (precision) and the fraction of predicted native contacts (sensitivity). In particular, since coevolution in RNA
	is expected to be related to isostericity \cite{leontis2002non,stombaugh2009frequency}, we only considered base pairing and excluded other base-backbone or backbone-backbone
	contacts.

	Our results show that approximately 40\% of the total native contacts can be predicted by this procedure. A large fraction of the predicted contacts are secondary structure contacts or pseudoknotted helices. However, in most of the analyzed structures, at least one tertiary contact is correctly predicted. In addition, the number of false positives is very small ($\approx 10\%$ of the predicted contacts). In many cases, false positives are just labeled so by our decision to exclude stacking interactions from the true contacts. In other cases, false
	positives are a consequence of an erroneous alignment of some of the sequences.
Some false positives are genuinely caused by numerical noises or by the assumptions behind the Potts model.
We notice that in principle the detrimental effect of false positives on the
	accuracy of structure prediction might be mitigated by using approaches where contacts that are not compatible with the predicted structure
	are discarded iteratively \cite{weinreb20163d}.
	As a general consideration, it must be kept in mind that strong couplings as predicted by DCA are a signature of co-evolutionary pressure
	but not necessarily of spatial proximity.
For instance, functionally related elements that are far from each other in space might exhibit coevolution.
	Relationships of this kind could in principle decrease the precision of the method in predicting contacts. In principle, highly conserved residues carry a limited amount of information and could thus reduce the sensitivity of the method, although in practice we never observed
	a very high conservation in the analyzed bacterial sequences. Eukariotic sequences might be more sensible to this issue, as it can be
        seen by the lower performance of the method when applied to spliceosomal RNA.
	
	Importantly, we developed a rigorous manner to establish a threshold for contact prediction. In particular, once a figure of
	merit capable to take into account both the method precision and sensitivity has been defined, an optimal threshold can be found on a specific training set. We here used the Mathews correlation coefficient, that corresponds to the interaction network fidelity \cite{parisien2009new} widely used in the RNA structure-prediction community \cite{miao2017rna}.
	The resulting thresholds are different depending on the used method (mutual information vs the tested DCA methods), but are
	transferable across different RNA families as illustrated by our cross-validation analysis.

It is important to observe that RNA molecules often display dynamics (i.e. coexistence of multiple structures) related to function, and that perhaps riboswitches
are the paradigmatic example where multiple structures are required for function.
For instance, some of the false positives might correspond to true contacts in an alternative, biologically functional structure
(e.g., on and off state of the riboswitch).
This fact might affect the results of the comparison reported here.
Nevertheless, we believe that high resolution X-ray structures still represent the best proxy for the correct solution structure and as such they should be used
for a critical assessment. Without having an experimentally determined ensemble, it appears difficult to assume that the observed false positives
are, by chance, important contacts in alternative structures.

A crucial finding is that the here introduced stochastic solution of the inverse problem (Boltzmann learning) is feasible on these systems
and outperforms the other DCA approaches.
The resulting Potts models were shown to reproduce correctly the empirical frequencies from the MSA.
Whereas the fact that the mean-field approach provides an approximate solution is well-known
\cite{nguyen2017inverse,cocco2018inverse}, no such comparison has been reported on RNA DCA yet.
In addition, we show that, although it is supposed to be capable to infer correct couplings at least in the limit of a large
number of sequences, also the pseudo-likelihood approximation is not capable to reproduce the correct frequencies with the employed datasets.
This fact was recently observed for protein systems \cite{figliuzzi2018pairwise}, where it was also observed that in spite of this
disagreement the contact predictions obtained with the pseudo-likelihood approximation are of quality similar to those obtained with Boltzmann learning DCA.

The overall improvement in the accuracy of the predictions, as measured by the MCC, when passing from state-of-the-art pseudo-likelihood DCA to
Boltzmann-learning DCA is comparable to the one observed when passing from mean-field DCA to pseudo-likelihood DCA,
which has been already shown to improve the quality of 3D structure prediction \cite{de2015direct}.
	It is worth saying that the extra cost of the Boltzmann learning procedure is significant if one wants to characterize a large number of families.
The required times for all the tested covariance methods scale roughly as the number of nucleotides squared and are listed in table \ref{table} for the largest and smallest molecules in the data set.
If we also include
	the cost of a later 3D structure prediction and refinement, we consider the extra cost of Boltzmann learning to be absolutely worth. We believe that the fast Boltzmann learning procedure introduced here based on a stochastic gradient descent could be
	fruitfully used in protein systems as well.
	We also notice that the stochastic procedure used here is closely related to similar techniques used in the molecular dynamics community in
	order to enforce preassigned distributions in the generation of molecular structures \cite{cesari2018using,valsson2014variational,white2014efficient,cesari2016combining}.
	We chose here to use the simplest possible optimization algorithm, but more advanced procedures might make the Boltzmann learning
	approach even faster.

		\begin{table}
			\caption{Computational time for the smallest and largest system investigated. Machine hardware architecture: Intel E5-2620, 12 physical cores. Operating system: GNU/Linux.
				Mutual information, MF-DCA, and BL-DCA predictions were done using in house code. R-scape predictions were done using R-scape 1.2.3. PL-DCA were done using plmDCA\_asymmetric\_v2 code available on GitHub.}
			\begin{center}
				\label{table}
				\renewcommand{\arraystretch}{1.3}
				\begin{tabular} {c|c|c}
					
					\bf{Method} & \bf{3DOU (largest)} & \bf{3VRS (smallest)} 
					\\\hline
					Boltzmann learning DCA & 220 min& 20 min  \\
					Pseudo-likelihood DCA& 3 min & 30 sec  \\
					R-scape & 33 sec & 9 sec  \\
					Mean field DCA & 22 sec & 4 sec \\
					Mutual Information & 15 sec & 3 sec \\

				\end{tabular}
			\end{center}
		\end{table}

We also tested the state-of-the-art pseudo-likelihood maximization approach, which
is faster than the Boltzmann learning approach but, on the tested dataset, provides results of slightly inferior quality.
Interestingly, the relatively good contact predictions obtained using pseudo-likelihood DCA are not paralleled by correct frequencies in the reconstructed Potts model.
Similar results were obtained decreasing the regularization term usually employed in pseudo-likelihood DCA.
This effect is likely due to the finite number of available sequences.
A more important practical issue is that the optimal threshold used for contact predictions resulted less transferable across different families
in pseudo-likelihood DCA when compared with Boltzmann learning DCA. This suggests that choosing a cutoff that can single out true contacts might be more
difficult in this method. 

The impact on contact prediction of other sometime overlooked choices (reweighting and APC correction) has also been assessed. Our results show that these
choices lead to negligible or minor improvements to all the methods.

	Finally, we show that
	the alignment procedure used to prepare the MSA has a significant impact on the accuracy of the prediction. Interestingly, the \textit{Infernal} algorithm, that
	is based on a previous prediction of the secondary structure, performs significantly better than the \textit{ClustalW} algorithm.
Whereas this effect is somewhat expected, we are not aware of similar assessments done on DCA methods.
	We observe that the couplings obtained with the present approach might be used to further refine the multiple sequence alignments.

	In conclusion, the direct coupling analysis method was assessed on a number of RNA families.
	We found that, in spite of the intrinsic approximations, this procedure is able to reliably predict a number of contacts in 
	RNA molecules with known three-dimensional structure. Among the tested methods, the Boltzmann learning approach is the one that 
	allows to simultaneously maximize accuracy and precision.
	In perspective, we foresee the possibility to explicitly use information about the isosteric RNA families \cite{leontis2002non,stombaugh2009frequency} or include three-body terms \cite{schmidt2017three} in order
	to further improve the accuracy of the predictions.
	Ultimately, we suggest the direct coupling analysis performed through the Boltzmann learning as the best available tool to enhance RNA structure prediction
        for systems of up to a few hundreds nucleotides, taking advantage of only homologous sequences information.

\section{Acknowledgements}

Daniele Granata, Anna M.~Pyle, Petr \v{S}ulc and Eric Westhof are acknowledged for reading our manuscript and providing enlightening suggestions.

\end{document}